# Government Investments and Entrepreneurship

João Ricardo Faria
Department of Economics, Florida Atlantic University, USA

Laudo Ogura
Department of Economics, Grand Valley State University, USA

Mauricio Prado
Department of Economics, Copenhagen Business School, Denmark

Christopher J. Boudreaux
Department of Economics, Florida Atlantic University, USA
cboudreaux@fau.edu

**Abstract:** How can governments attract entrepreneurs and their businesses? The view that new business creation grows with the optimal level of government investments remains appealing to policymakers. In contrast with this active approach, we build a model where governments may adopt a passive approach to stimulating business creation. The insights from this model suggest new business creation depends positively on factors beyond government investments—attracting high-skilled migrants to the region and lower property prices, taxes, and fines on firms in the informal sector. These findings suggest whether entrepreneurs generate business creation in the region does not only depend on government investments. It also depends on location and skilled migration. Our model also provides methodological implications—the relationship between government investments and new business creation is endogenously determined, so unless adjustments are made, econometric estimates will be biased and inconsistent. We conclude with policy and managerial implications.

**Plain English Summary.** Governments can attract entrepreneurs and their businesses by offering incentives such as lower property prices, taxes, and fines on firms in the informal sector, as well as by encouraging skilled migration to the region. Thus, a policy implication is that the government can create a favorable environment for business creation, as opposed to solely relying on government investments.

## 1. Introduction

In 2017, Amazon received 238 proposals from North American cities attempting to attract Amazon's second corporate headquarters location.[1] This attracted international attention—one of the world's largest companies requested and received proposals from these cities to offset at least part of its proposed $5 billion investment. This is not the first such case. In 1976, Volkswagen decided to locate their first U.S. manufacturing plant in Pennsylvania after receiving multiple bids from several states (Slattery, 2020), with many other cases following.[2] In 1993, Alabama persuaded Mercedes-Benz to build its first U.S. plant in the town of Vance. How? By offering an incentive package worth $253 million or about $169,000 for every job Mercedes-Benz promised.[3] Another well-known example is BMW's plant in Spartanburg, South Carolina[4] (Boudreaux et al., 2012; Greenstone and Moretti, 2003).[5]

This competition between regional governments to attract entrepreneurs[6]—the coordinators of the factors of production (Say, 1828)—is hardly surprising given that entrepreneurship drives economic growth (Acs et al., 2016; Acs and Szerb, 2007; Holcombe, 1998; Minniti, 2008). Studies have shown that countries with more entrepreneurs grow faster (Acs, 2006; Audretsch et al., 2006; Murphy et al., 1991), especially when entrepreneurs are high-growth firms or "gazelles" (Acs and Mueller, 2008; Coad and Srhoj, 2020; Henrekson and Johansson, 2010).[7] Accordingly, government officials have implemented policies to support entrepreneurship (Acs et al., 2016;

---

[1] https://www.brookings.edu/blog/the-avenue/2017/12/01/u-s-cities-compete-for-amazon-will-tax-incentives-deliver-hq2/
[2] Good Jobs First. (2022). Subsidy Tracker. https://www.goodjobsfirst.org/subsidy-tracker
[3] https://www.wsj.com/articles/SB1017784548687093560
[4] https://www.latimes.com/archives/la-xpm-1992-06-23-fi-922-story.html
[5] Greenstone and Moretti, 2003 studied the location of million-dollar business investments announced in the 1982-1993 period and provided an economic analysis of the value the local region of attracting large firms by weighing the benefits of employment and property value gains against the costs of incentive packages.
[6] There are many different conceptualizations of entrepreneurs. Parker (2018) provides an excellent summary. We follow the advice of Welter et al. (2017) and refrain from adopting a narrow conceptualization of entrepreneurship.
[7] See Packard and Bylund (2018) for a criticism on the focus of this type of entrepreneurship.

Lerner, 2010; Mason and Brown, 2013; Shane, 2009). These support policies sometimes take a relatively "hands off" approach such as fostering a favorable environment to encourage entrepreneurship (Minniti, 2008; Obaji and Olugu, 2014; Urbano et al., 2019). In other cases, such as the examples above, governments take a more active approach. This often entails providing tax incentives, subsidizing R&D, and "picking winners" (Autio and Rannikko, 2016; Buffart et al., 2020; Mazzucato, 2018).

Despite the importance of these studies, we still know little about the effectiveness of such policies. On the one hand, some have argued the government has an important role to fill in promoting entrepreneurship because of its ability to foster job creation (Goodman, et al., 1992), innovation (Michael and Pearce, 2009), and benefits to the local economy (Greenstone and Moretti, 2003). Governments can also expand their approach from targeting specific firms to making the region more inviting to businesses to attract many firms. David Audretsch refers to this as the "strategic management of places" (Audretsch, 2003). Governments accomplish the former by providing tax incentives and subsidizing R&D and the latter by increasing expenditures on beautification and infrastructure projects (Audretsch et al., 2015; Bennett, 2019a; Van De Ven, 1993). On the other hand, although well-intended, scholars argue this approach has had little effect on entrepreneurship activity, and the resources can be put to better use (Acs et al., 2016, Lee, 2018; Lerner, 2010; Shane, 2009). In the worst-case scenario, it is possible that government policies, like socially-inefficient regulation, might inhibit formal entrepreneurship (Boudreaux et al., 2018; Prado, 2011). Moreover, Bylund (2016) disputes that there is such a thing as "socially-efficient" regulation.

In light of the above, the purpose of our paper is to examine two approaches local governments might adopt to support entrepreneurship. Although a consensus has emerged that institutional factors are the "driving conditions for entrepreneurship" (Aparicio et al., 2021; Urbano et al., 2019, p. 24), there is less agreement on the efficacy of the active approach for government investments in entrepreneurship (Lerner, 2010; Shane, 2009)[8]. For instance, consider the policies heralded by Mazzucato (2018, 2015) to promote a more active government as the "entrepreneurial state" and the criticisms of this view (Wennberg and Sandström, 2022). Thus, there remains a limited understanding and agreement of how and when government investments in entrepreneurship can encourage and support entrepreneurship. For instance, Bennett (2019a) found that private investments help encourage job creation while government investments encourage job destruction. The results show that economic freedom is positively associated with firm and job creation, but it has no effect on firm and job destruction. One implication is that all investments are not created equal, so it is not enough to propose that more investments will necessarily lead to greater entrepreneurship and growth. Although this is a good first step, studies show the relationship between the private and public sector varies greatly between countries (Gwartney et al., 2019) as well as the relationship between entrepreneurship and economic growth varies between countries (Bjørnskov and Foss, 2016; Praag and Versloot, 2007; Wennekers and Thurik, 1999). Bennett's (2019a) study is based only on U.S. data and the conclusion likely differs based on a different context or composition of the private and public sector.

To answer these questions, we construct a model in which government investments in infrastructure may attract new business creation into the region in the formal and informal sectors. There is evidence that under developing countries tend to have more entrepreneurial activity than developed ones (Bosma et al., 2020), and the share of entrepreneurs is disproportionately in the informal sector in developing countries (De Soto, 2000). Importantly, and in contrast to prior studies (Mendicino and Prado, 2014; Prado, 2011), we examine two institutional scenarios—one in which the government is the follower and another where the government is the leader. The former is consistent with the hands-off approach to government support policy and the latter is consistent with the active approach prevalent in the literature. The active approach, where government is the leader, operates under the assumption the state ought to drive economic growth. In contrast, the findings from our model suggests the optimal location of entrepreneurs is independent of government expenditures in infrastructure under the hands-off approach. Instead, the location decision depends on whether policies attracting high-skilled migration outweigh rising property prices and taxes.

---

[8] Although we discuss government infrastructure investments, government investments can be quite broad. For example, according to Lerner (2010), Singapore implemented specific government policies such as: "The provision of public funds for venture investors seeking to locate in the city-state; Subsidies for firms in targeted technologies; Encouragement of potential entrepreneurs and mentoring for fledgling ventures; Subsidies for leading biotechnology researchers to move their laboratories to Singapore; Awards for failed entrepreneurs (with a desire to encourage risk-taking)."

Our study makes several contributions to the literature for both theory and practice. First, we expand the current models of government investments and entrepreneurship to consider both the passive and the active approaches to government support policy. For instance, Faria et al. (2021) show that a business' optimal location is a function of government expenditures in infrastructure. This study, however, does not consider the heterogeneity in government support policies. Moreover, although there is evidence suggesting that private investments might be more effective than government investments in entrepreneurship (Bennett, 2019a, 2019b), we need more research to identify how and when government investments are effective (Wennberg and Sandström, 2022). For example, Audretsch et al. (2015) and Bennett (2019a) both find heterogenous effects of infrastructure investments. Audretsch et al. (2015) find that broadband investments are more effective than highway and roads and Bennett (2019a) finds that private infrastructure investments are more effective than public infrastructure investments. Furthermore, there have been calls for a diversity of methodological approaches to investigate these issues (Parker, 2020).

Second, we heed calls for the use of advanced methods to the multidisciplinary entrepreneurship literature (Parker, 2020). Specifically, we study a general equilibrium model, i.e., a model with several sectors interlinked, which offers a first step in the direction of identifying how and when government investments are effective. Mathematical theoretical models can offer valuable insights, counter-intuitive results, and are based on a core set of assumptions. Economic theory is necessary to formulate testable hypothesis, i.e., hypothesis which can be falsified (Popper, 1934). Thus, our model provides a more complete theoretical formulation over earlier studies (Bennett, 2019a, 2019b) with the advantage that we develop more testable hypotheses. Bennett's empirical tests can be seen as one possible empirical implementation of our paper.

Third, our study contributes to the empirical econometric literature on government policy and entrepreneurship. Our model reveals that the relationship between government investments and new business creation is endogenously determined, so unless adjustments are made, it will render biased and inconsistent econometric estimates.

Lastly, our study has important implications for public policy. There are mixed findings in the literature for the role of active programs to facilitate entrepreneurship because policies like tax incentives and subsidies tend to encourage the typical start-up rather than high-growth gazelles (Lerner, 2010; Shane, 2009). Under the assumption that policy-makers want entrepreneurship for the potential of employment and job creation, policies focusing on the typical start-up will be unlikely to generate the results desired. Our model suggests governments can encourage entrepreneurship by adopting a passive approach, thereby suggesting policymakers might focus on place-based policies (Audretsch, 2003) to attract start-ups rather than targeting specific start-ups.

## 2. Literature Review - Infrastructure and Government

New infrastructure investments help facilitate the flow of capital, goods, ideas and people, and raise productivity (Bucovetsky, 2005; Taylor, 1992). In a paper on Germany, Audretsch et al. (2015) find that regional start-up activity is positively and significantly associated with physical infrastructure. Bennett (2019a) presents a model of infrastructure investments and uses annual U.S. state-level data to show that private infrastructure investment is positively and significantly associated with creation of businesses and jobs, while public infrastructure investments are associated with the destruction of businesses and jobs. Examples of private infrastructure investments include buildings and structures for lodging, offices, private public safety, non-railroad transportation, highway and street, sewage and waste disposal, water supply, and conservation and development. Private infrastructure investments exclude projects in the power, communication, and railroad sectors, but public infrastructure investments do not (Bennett, 2019a).[9]

Market failures in the private sector and the acknowledgement of the benefits of entrepreneurship for the economy have alerted policymakers to embrace entrepreneurship through public policy (Acs et al., 2016; Acs and Szerb, 2007; Henrekson and Stenkula, 2010; Holtz-Eakin, 2000; Padilla-Pérez and Gaudin, 2014). Entrepreneurship policy is a government intervention in the market designed to increase either the quantity or the quality of entrepreneurship (Block et al., 2018; Fritsch, 2008; Parker, 2018). The presence of market failures that inhibit individuals from launching innovative new ventures are used to justify government interventions intended to encourage entrepreneurship (Acs et al., 2016; Ács et al., 2014; Padilla-Pérez and Gaudin, 2014).

---

[9]Although Bennett (2019b) excludes projects in the power, communication, and railroad sectors from private infrastructure investments, there are historical examples of private companies investing in these sectors. For example, James J. Hill built the Great Northern Railway from St. Paul to Seattle with no U.S. government aid (Folsom, 1991).

Many countries attempt to increase the number of entrepreneurs and to improve their performance by promoting educational and training programs along with reducing administrative barriers and providing start-up subsidies (Andersson and Wadensjö, 2007; Coomes et al., 2013; Fredriksson, 2020). Such efforts have shown significant payoffs in terms of economic growth and job creation (Henrekson and Sanandaji, 2011; Parker, 2018; Wennekers and Thurik, 1999)

Similarly, policymakers have considered granting the government a larger role in policy at local and regional levels of government (Lucas and Boudreaux, 2020). Government policies aimed at attracting multinational enterprises generating positive regional spillovers have been examined in regards to location decision. In a study of foreign direct investment (FDI) in Ireland, Barrios et al. (2006) find that regional policies tend to attract low-tech firms to disadvantaged areas. Devereux et al. (2007) observe that discretionary government grants have a small effect in attracting FDI to specific areas, and Crozet et al. (2004) uncover scant evidence that regional policies have a beneficial impact on FDI. Faria (2016) shows that the most efficient public policies to attract foreign firms to a region are the provision of logistical infrastructure, creation and/or nurturing of pro-market institutions, non-discretionary policies, reduction of statutory tax rate, and incentives to universities and research centers.[10] Active government support policies like transfer payments, loans, subsidies, regulatory exceptions, and tax benefits for starting a business generally fail to generate sustainable growth and to solve market failures (Acs et al., 2016; Shane, 2009).

As pointed out by Hayter (1997) there are three approaches for firms' location decisions. The neoclassical approach (Faria, 2016; Grimes, 2000; Ouwersloot and Rietveld, 2000) focuses on strategies for maximizing profits and minimizing costs, including how economic clustering helps productivity of local firms (Abdel-Rahman, 1988; Feldman et al., 2005; Glaeser et al., 2010). The institutional approach (Arauzo-Carod and Viladecans-Marsal, 2009; Galbraith, 1985) considers how companies locate taking into account not only distance to consumers, suppliers and other companies, but also the institutional surroundings given by commercial associations, regional systems, and the local government. The behavioral approach (Dahl and Sorenson, 2012; Sabat and Pilewicz, 2019; Sorenson, 2018) considers informational issues and uncertainty. Our model combines these elements since optimal location of firms depends on profit maximization, government policies, and whether or not the firm operates in a formal or informal sector, reflecting the degree of business uncertainty.

---

[10] This is in line with Thurik and Wennekers (2004) and Acs et al. (2016), which support government policy that targets skills supply through education, knowledge transfer, worker mobility, and the ability to start new firms.

## 3. The Model

Our model is based on an analogy to Schumpeter's creative destruction hypothesis (Schumpeter, 1934) to explain how regional-level institutional differences and government policies influence entrepreneurial activity related to the creation and destruction of firms. For instance, countries and states whose institutions better reflect principles of economic freedom have higher levels of entrepreneurship (Bennett, 2021; Bjørnskov and Foss, 2016; Boudreaux et al., 2019, 2022; Boudreaux and Nikolaev, 2019; Coomes et al., 2013; Nikolaev et al., 2018; Urbano et al., 2019). Business creation $B$ is an increasing function of government creative destruction $s$, capital $K$, and location $L$:

$$B_t = B(s_t, K_t, L_t) \quad (1)$$

where $B$ grows with all three arguments in (1).

The government invests in infrastructure $G$ in the region (roads, water, electricity, etc.). Some local businesses thrive because of these investments, while others are hurt by new competition or may find it more profitable to move to other locations[11]. This government-induced creative destruction of firms (Bennett, 2019a) is expressed as the difference between business creation $C(G)$ [led by the government] and business destruction $D(G)$:

$$s_t = C(G_t) - D(G_t) \quad (2)$$

The location variable $L$ is an index that combines the real estate stock and the relative value of its location [e.g., whether it is closer to demand or input suppliers]. Entrepreneurs pay attention to infrastructure investments that create business value, thus trying to locate in the region benefited by government creative destruction:

$$L_{t+1} - L_t = \varphi L_t \left( s_t - \frac{L_t}{Q} \right) \quad (3)$$

where $\varphi$ is a positive parameter and $Q$ is the carrying capacity of the environment in sustaining the expansion of $L$.[12]

In Equation 1, location $L$ helps business creation directly as there are agglomeration economies, that is, the productivity advantages from firm clustering due to the availability of resources like venture capital and skilled labor, among other reasons (Glaeser et al., 2010; Luo et al., 2020).

Firms invest $I$ in location $L$ and capital $K$. As a consequence, the accumulation equation for capital, $K_{t+1} - K_t$, is given by the difference between investment $I$ and expenditures in location $pL$, i.e., capital accumulation is made with resources that are not spent in properties:

---

[11] An example of the type of infrastructure investments that we have in mind is a government investment in rural high-speed broadband (Cumming and Johan, 2010) which is quite different than some other type of investment, like neighborhood beautification, in terms of how it affects entrepreneurship. Because the focus here is on job creation and destruction, we ignore the effects of government policy on low growth entrepreneurship, which deserves to be studied in a framework where its contributions to society are accounted appropriately, as advocated by Welter et al. (2017).

[12] The $Q$ in the model is a parameter that represents the spatial restrictions in the regional economy due to distance, negative agglomeration externalities, and/or regulations. The $Q$ is assumed exogenous, being determined by regional idiosyncrasies like topography and land use regulations, which are typically disconnected from the government levels or entities that provide business incentives.

$$I_t = K_{t+1} - K_t + pL_t \rightarrow K_{t+1} - K_t = I_t - pL_t \quad (4)$$

where $p$ is the property price of locating in the region. Remark: the purchase price of capital is constant equal to 1 and the depreciation rate of capital is assumed to be zero.

There are two sectors in the economy: formal ($f$) and informal ($i$), and $\alpha_j$ measures the share of firms in each sector, where $j = \{i,f\}$. On this regard we assume the formal sector is larger, $\alpha_f > 0.5$, and:

$$\alpha_f + \alpha_i = 1 \quad (5)$$

Firms face adjustment costs associated with new investments in capital and location. These are impacted by informality as well. Firms in the formal sector pay a tax $\tau$, while firms in the informal sector do not pay a tax but instead pay a fine $e$ when caught. Thus, adjustment costs are $c(e, \tau, I_t)$.

In line with the "family and friends effect" literature (Morgan et al., 2018) on immigration, we assume that high skilled immigrants $M$ migrate legally and create businesses that facilitate the immigration of unskilled workers $U$ [legal or illegal]. Being high skilled and legal, $M$ create businesses in the formal sector. By the same token, unskilled immigrants may be employed in formal or informal sectors.

$$U_{t+1} - U_t = z\alpha_f M_t - (v - \alpha_i)U_t \quad (6)$$

where the first term in the right-hand side, $z\alpha_f M_t$, captures business creation in the formal sector of high skilled migrants. Note that legal immigration $M$ is exogenous. In the last term in the right-hand side, $v > \alpha_i$ is a parameter reflecting the identification and deportation of illegal immigrants, who predominantly would have worked in the informal sector.

The aggregate production function in the economy $F(K_t, N_t, L_t, G_t)$ is defined as follows:

$$F(K_t, N_t, L_t, G_t) = \left(1 + L_t x G_t^{\frac{1}{x}}\right) K_t^a N_t^{1-a} \quad (7)$$

where $0<a<1$, $L_t x G_t^{\frac{1}{x}}$ is a level of productivity-enhancing public goods[13] G interacted with location L, and $N_t = g U_t + (1-g) u_t$ is unskilled labor, which is a convex combination of immigrants ($U$) and domestic workers ($u$). The parameter $g$ is the proportion of immigrants in the labor force.

In what follows we solve the model for two different types of government action: 1) the government is the follower and 2) the government is the leader. The government is market friendly when it is the follower. It is defined by the market determination of the optimal level of government investments $G$. In our framework, this means $G$ is endogenously determined by the entrepreneurial decisions and the government acts only to acquiesce the demands of the private sector in establishing $G$. When the government acts as leader, it intends to direct and stimulate private investment and business creation according to its own plans meaning the government maximizes Equation (2) with respect to $G$, independently from entrepreneurs' interests.

### 3.1. Government as a Follower Equilibrium [Free Market or Tobin Equilibrium]

In the "follower government" equilibrium, the optimal level of government investments $G$ is endogenously determined by entrepreneurial decisions and the government acts only to satisfy the level of $G$ demanded by the private sector.

This equilibrium is obtained by solving the representative entrepreneur problem. The representative entrepreneur maximizes the present value of profits net of adjustment costs:

$$\max V = \sum \beta^t [F(K_t, N_t, L_t, G_t) - w_t N_t - I_t - c(e, \tau, I_t)] \quad (8)$$

subject to Equations (3)-(7). Where $\beta = 1/(1+r)$ is the discount factor and $c(e, \tau, I_t)$ are the adjustment costs associated with choosing and combining capital and location affected by taxes or fines whether the business is in the formal or informal sectors. We assume $c_I(e, \tau, I_t) > 0$ ; $c_{II}(e, \tau, I_t) > 0$.

---

[13] We assume decreasing returns in the public-good investment technology, such that $x>1$, as in Acemoglu (2005).

The Lagrangian corresponding to the representative entrepreneur problem is:

$$\sum \beta^t \left\{ \left(1 + L_t x G_t^{\frac{1}{x}}\right) K_t^a N_t^{1-a} - w_t N_t - I_t - c(e, \tau, I_t) + \mu_t[I_t - pL_t + K_t - K_{t+1}] + $$

$$\vartheta_t \left[ \varphi L_t \left( s_t - \frac{L_t}{Q} \right) - L_{t+1} + L_t \right] \right\} \ (9)$$

The choice variables are: $I_t, K_{t+1}, L_{t+1}, N_t$. In the Appendix we derive the first order conditions [FOC] with respect to the choice variables and the free market equilibrium. Table 1 reports the comparative static analysis, which shows how the parameters and exogenous variables affect the endogenous variables of the Tobin's equilibrium.

### 3.2. Government as a Leader Equilibrium [Dirigist Government]

As defined above, a leading government maximizes the creative destruction of its investments, which implies in maximizing Equation (2) with respect to $G$, independently from entrepreneurs' interests, which yields in the steady state:

$$Max \ s \rightarrow C_G(\hat{G}) = D_G(\hat{G}) \quad (10)$$

Condition (10) equates the marginal benefits of government investments to marginal costs. The hat over the variable defines the optimum government investment $\hat{G}$ of the active government equilibrium.

Considering the government has the monopoly of public goods provision, condition (10) which yields a maximum creative destruction, implies that $\hat{s}$ has to be positive,

$$\hat{s} = C(\hat{G}) - D(\hat{G}) > 0 \ (11)$$

Considering that Equations (10) and (11) are given to the entrepreneur, solving the problem in the steady-state implies

$$\varphi L \left( s - \frac{L}{Q} \right) = 0 \quad (12)$$

According to condition (12) we have two choices: 1) Either $s = \frac{L}{Q}$ and Equation (11) is equal to the free market equilibrium condition and there is no difference whether the government is a leader or a follower, i.e., $G^* = \hat{G}$, or 2) $\varphi = 0$, and the optimal location $L_{t+1} = L_t = \hat{L}$ is constant *ab ovo* and the fact that $s \neq \frac{L}{Q}$ does not matter at all, implying we can consider $\hat{L}$ as exogenous.

A constant and exogenous $\hat{L}$ implies that Equation (4) determines optimal investment as:

$$\hat{I} = p\hat{L} \quad (13)$$

An important remark has to be made here, since $\hat{L}$ is constant, $\hat{I}$ has to satisfy (13) since the beginning and, therefore, $\hat{K}$ is also constant since the beginning in order to obtain a steady state. The main implication is therefore when the government is the leader, there is no growth either in location or in capital. An example is cleaning and beautification efforts that do not change the nature of the businesses in the location benefited by those efforts.

Then given $M$ by Equation (6) we determine the optimal level of unskilled immigration $\hat{U}$:

$$\hat{U} = \frac{z\alpha_f M}{(\nu - \alpha_i)} \quad (14)$$

### 3.3. Comparative Statics Analysis

Table 1 summarizes the comparative statics analysis of the Tobin's Equilibrium. We first focus on the roles of immigration, informality, and government infrastructure in business creation.

First, we note properties prices $p$, the tax paid by the formal sector $\tau$, and the enforcement fine on informal firms $e$ all negatively affect the location index $L$. Similar to location, $p$, $e$, and $\tau$ all negatively affect the immigration of low skilled workers $U$. In addition, identification and deportation of illegal immigrants $v$, and the share of firms in the formal sector $\alpha_f$ negatively affect immigration as well, while the share of firms in the informal sector $\alpha_i$ and business creation in the formal sector of high skilled migrants ($z$ and $M$) affect immigration positively.

**TABLE 1 COMPARATIVE STATICS OF TOBIN'S EQUILIBRIUM**

|   | $a$ | $p$ | $v$ | $x$ | $z$ | $E$ | $\tau$ | $\alpha_i$ | $\alpha_f$ | $\varphi$ | $Q$ | $M$ | $r$ | $G$ |
|---|---|---|---|---|---|---|---|---|---|---|---|---|---|---|
| $I$ | 0 | 0 | 0 | 0 | 0 | - | - | 0 | 0 | 0 | 0 | 0 | 0 | 0 |
| $L$ | 0 | - | 0 | 0 | 0 | - | - | 0 | 0 | 0 | 0 | 0 | 0 | 0 |
| $s$ | 0 | - | 0 | 0 | 0 | - | - | 0 | 0 | 0 | - | 0 | 0 | 0 |
| $U$ | 0 | - | - | 0 | + | - | - | + | - | 0 | 0 | + | 0 | 0 |
| $G$ | 0 | - | 0 | 0 | 0 | - | - | 0 | 0 | 0 | - | 0 | 0 | 0 |
| $N$ | 0 | - | - | 0 | + | - | - | + | - | 0 | 0 | + | 0 | +/- |
| $K$ | + | - | - | +/- | + | - | - | + | - | 0 | - | + | - | +/- |
| $w$ | + | +/- | +/- | +/- | +/- | +/- | +/- | +/- | +/- | 0 | - | +/- | - | +/- |
| $\vartheta$ | + | - | - | +/- | + | - | - | + | - | + | +/- | + | - | +/- |
| $B$ | + | - | - | +/- | + | - | - | + | - | 0 | - | + | - | +/- |

*Note* +/- stands for ambiguous, where effect depends on other values.

The equilibrium business creation $B$ is negatively impacted by $p$, $v$, $e$, $\tau$, $\alpha_f$, Q (the carrying capacity of the environment in sustaining the expansion of $L$) and the interest rate $r$. Business creation is positively affected by $z$, $M$ and by the capital share in the production function $a$. The degree of productivity-enhancing public goods interacted with location $x$ and the proportion of immigrants in the labor force $g$ have both ambiguous impacts on $B$.

In short, holding other factors constant, immigration helps business creation as it enlarges the labor force[14], while informality also helps business creation as it stimulates immigration of low skilled workers.[15] In the free market equilibrium, government infrastructure and business creation are determined simultaneously, so we cannot say that G affects B.

---

[14] The reverse is also likely true—business creation tends to attract more workers and immigrants. However, in our model, *M* is exogenously given, so the only causality possible in our model is from immigration to business creation. We thank one anonymous reviewer for pointing this out.

[15] Bandyopadhyay and Pinto (2017) and Ogura (2018) find similar results in regards to the effect of immigration and informality on production.

As per the government as a leader equilibrium, the comparative statics is simpler since the location index $L$ is constant at $\hat{L}$. Immigration $U$ is given by Equation (14). Thus, $U$ grows with $z$, $M$, $\alpha_i$ and $\alpha_f$, and decreases with $v$. Last, but not least, business creation $B$ depends on two constants $\hat{K}$, $\hat{L}$, and grows with $\hat{s}$ which is determined by (12). Therefore, ultimately a growth in the optimal level of government investments $\hat{G}$ increases business creation.

### 3.3 Graphic Exposition of the Model

In what follows we present our location theory in graphs that help to expose its dynamics and equilibria. Figure 1 depicts Equation (3) that can be rewritten as:

$$L_{t+1} = L_t \left( \varphi s_t + 1 - \varphi \frac{L_t}{Q} \right) \quad (3')$$

Notice that from Equation (3') in the steady state equilibrium, $L_{t+1} = L_t$, we have multiple equilibria, $L_0, L_1, L_2$ , where $L_1$ is unstable and $L_0, L_2$ are stable.

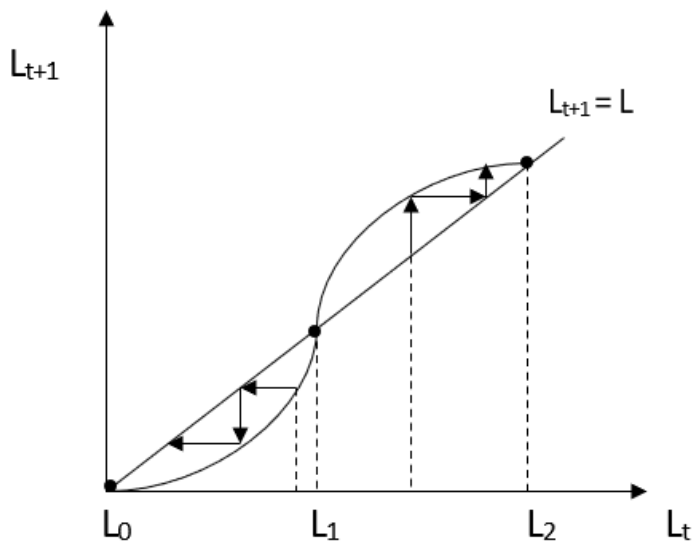

**Figure 1**. Multiple Equilibria and Stability

In the case of the government as the leader, as we saw $\hat{L}$ is exogenous. This implies that we must ignore the dynamics in Figure 1 and we end up with a graph like Figure 2 instead. In Figure 2, $\hat{L}$ can be at any point along $L_{t+1} = L_t$.

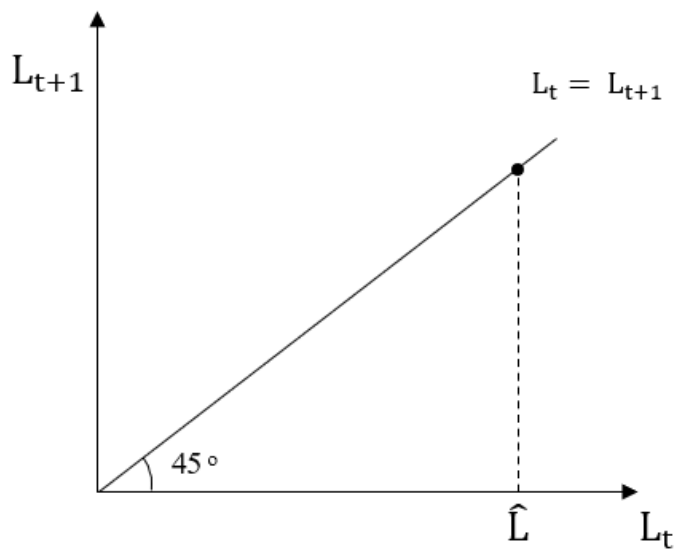

**Figure 2**. Government as Leader Equilibrium

In the free market Tobin equilibrium, the first order conditions determine optimum location from $L^* = I^*/p$. Consequently, Equation (3) yields optimum government creative destruction $s^* = L^*/Q$. The larger the L*, the larger the creative destruction: $L_2{}^* > L_1{}^* \rightarrow s_2{}^* > s_1{}^*$, and the growth rate increases as depicted in Figure 3.

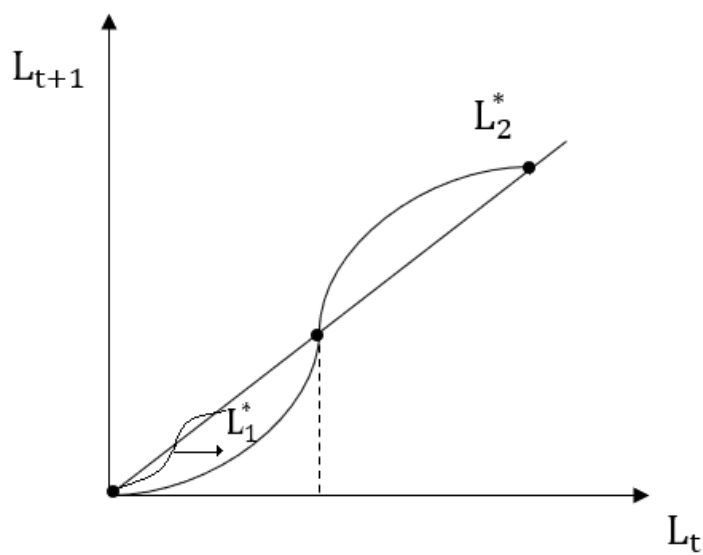

**Figure 3.** Tobin Equilibrium (Government as Follower)

## 4. Discussion

### 4.1. What is the best government role for business creation?

The purpose of this paper is to examine two approaches local governments might adopt to support entrepreneurship. Our model yields a number of important implications. In the free market equilibrium, in which the government is the follower, property prices, the tax paid by the formal sector, and the enforcement fine on informal firms all discourage locational investments and unskilled immigration. In addition, the identification and deportation of illegal immigrants and the share of firms in the formal sector hinders immigration. Importantly, these factors all discourage business creation[16]. Moreover, the carrying capacity of the environment in sustaining the expansion of a new location, and the interest rate also discourage business creation. In contrast, business creation and immigration encourage high skilled migration. When the government is a follower, the productivity-enhancing public goods (infrastructure) interacted with location and the proportion of immigrants in the labor force. This exerts an ambiguous impact on business creation.

---

[16] Even if infrastructure investments do not affect business creation directly, they might have an indirect effect through property prices, taxes, and immigration. As such, a possible extension of the model is to integrate G in these functions. However, this exercise would make exogenous variables like property prices, taxes, and immigration endogenous, so we miss their important impact in the model and in the policy prescriptions. Moreover, this extension changes the model entirely, yielding a different analysis and results. Thus, we leave this extension for future research.

In the case of the government as a leader, where creative destruction is maximized, business creation depends only on two equilibrium constants: capital and location. It grows with creative destruction; thus, the optimal level of government investments increases business creation.

In summary, governments can attract entrepreneurs and their businesses by offering incentives such as lower property prices, taxes, and fines on firms in the informal sector, as well as by encouraging skilled migration to the region.

One important question remains: which government stance is optimal for the creation of new businesses: a passive approach (i.e., free market equilibrium) or a more active approach (i.e., dirigist government)? To answer this question, note first that $\hat{L}$ and $\hat{K}$ in the government as a leader equilibrium are constant, which implies that they do not grow. By contrast, as the free-market equilibrium L* is obtained in the steady-state after periods of growth of the stock of capital, we must have $\hat{L} < L^*$ as well as $\hat{K} < K^*$. And since $\hat{s}$ is obtained from direct maximization we must have $\hat{s} > s^*$. As a consequence, we cannot *a priori*, without additional structure to the model, assess whether

$$B\left(\hat{s}, \hat{L}, \hat{K}\right) \overset{\geq}{\underset{<}{}} B(s^*, L^*, K^*) \quad (15)$$

Given inequality (15), the answer to the question of which is the type of equilibrium that leads to more business creation can only be answered empirically. The next subsection discusses some issues related to the empirical testing of these hypotheses.

### 4.2. *Empirical implications*

The literature (Audretsch et al., 2015; Bennett, 2019a) often estimates empirically the impact of government investments *G* on business creation *B* with a number of controls *Y*:

$$B = F(G, Y) \quad (16)$$

According to the results of our model, however, estimation of equations in line with Equation (16) are misspecified. Starting with the model of government as a leader, the empirical specification is simply:

$$B = F(G) \quad (17)$$

since $G$ is exogenously given through the government maximization problem and $B$ is an unequivocal function of it. So, the estimation of (17) is not consistent with using any controls $Y$.

Considering the Tobin equilibrium (i.e., the market model where $B$ and $G$ are endogenous variables), we have from the recursive causality of the model that $G$ is determined prior to $B$, so one is tempted to run $B$ as a function of $G$ and a number of parameters of the model:

$$B = F(G, p, v, e, \tau, \alpha_f, r, Q, z, M, x, a, g) \quad (18)$$

The problem with this approach is that $G$ being endogenous is a function itself of some of the parameters:

$$G = f(p, e, \tau, Q) \quad (19)$$

The endogeneity of $G$ makes the Equation (19) problematic, and the best econometric strategy is to run B as a function of the deep parameters of the model, i.e., without $G$:

$$B = F(p, v, e, \tau, \alpha_f, r, Q, z, M, x, a, g) \quad (20)$$

Therefore, this paper has important implications for the econometric strategy on how to assess the impact of government investments and creative destruction of business creation $B$. If $B$ is in fact a function of $G$, it is a unique function of it. Otherwise, $B$ is not a function of $G$ and it depends on the deep parameters of the model as in Eq. (20).

### 4.3. Policy implications

Our study has several implications to public policy. We find limited evidence for active support programs to facilitate entrepreneurship. As others have pointed out, this might be because government policies such as tax incentives and subsidies tend to encourage the average small business rather than high-growth gazelles (Lerner, 2010; Shane, 2009). Our model suggests governments can encourage entrepreneurship by adopting a passive approach to support policy.

But even in the passive approach, our model suggests the beneficial effect of attracting high-skilled migrants can be offset by the detrimental effect of rising property prices and taxes in the formal sector and fines in the informal sector. This is in line with our prior claim that the final impact of government on business creation is ambiguous and ultimately can only be set empirically. Therefore, in practice, government policies on infrastructure investment may be biased towards other social objectives, thus indirectly affecting business creation. For instance, in the model, immigration control was assumed exogenous, but the inflow of immigrants may negatively impact native people by increasing congestion of public resources and raising the competition in labor markets. As a result, government policies, including infrastructure investments, may end up biased towards hindering immigration (Bandyopadhyay and Pinto, 2017; Ogura, 2018). The model can be adjusted to consider more realistic policy environments, but the results do not change qualitatively.

### 4.4. Limitations and future research directions

Future research can speak to our claim that business creation and government investments are endogenously determined. Without special care, linear regression methods will report a biased and inconsistent parameter estimate of $\hat{G}$ on $B$. Studies will want to consider alternative methods such as instrumental variable methods, propensity or coarsened matching, Diff-in-Diff, and others to more precisely uncover the true parameter estimate in the population.

Our model also finds that a firm's location decision is negatively affected by property prices, taxes, and the fines levied on informal firms. In addition, the carrying capacity of the environment in sustaining entrepreneurship and the interest rate in the economy also inhibit business creation. Future research should empirically test these propositions in regards to how those factors interact with government entrepreneurship policies.

Last, but not least, there are two issues that deserve attention in future extensions of our theoretical model. First, governments tend to favor creating new programs while not getting rid of old programs (Cumming et al., 2017). This impacts the dynamics of government investments. Second, some government programs are complementary to one another, while others are substitutes (Cumming et al., 2018). These relationships also impact the dynamics and desirability of government incentives to entrepreneurship, deserving a deeper analysis.

## 5. Concluding Remarks

In this study, we construct a model to investigate whether government infrastructure investments support or inhibit the net creation and location of businesses. We examine two institutional scenarios, free market economies in which the government is the follower, and an alternative scenario where the government is the leader, reflecting a more socialized economy.

The paper yields important implications. In the free market equilibrium, properties prices, taxes paid by the formal sector, and enforcement fines on the firms in the informal sector all discourage locational investments and immigration. Immigration is also discouraged by the identification and deportation of illegal immigrants and the share of firms in the formal sector. Business creation is discouraged by those factors, the carrying capacity of the environment in sustaining the expansion of location, and the interest rate. In contrast, business creation and immigration are encouraged by high skilled migration. In the case of government as a follower, the productivity-enhancing public goods interacted with location and the proportion of immigrants in the labor force exert ambiguous impacts on business creation.

As per the government as a leader equilibrium, where creative destruction is maximized, the comparative statics are simpler since location is constant. Business creation depends on two constants in steady-state: capital and location, and grows with government creative destruction. Ultimately, growth in the optimal level of government investments increases business creation.

Last, but not least, our study offers a methodological critic of the usual estimations of the impact of government investments and creative destruction of business creation $B$. For the econometric strategy, we argue, that if $B$ is in fact a function of Government investments $G$, it is a unique function of it. Otherwise, $B$ is not a direct function of $G$ and it depends on the deep parameters of the model.